\documentclass[a4paper]{article}
\usepackage{graphicx}
\usepackage{amsmath,amssymb,amsfonts}
\newcommand{\be}{\begin{equation}}
\newcommand{\ee}{\end{equation}}
\newcommand{\tr}{{\rm Tr}}

\begin{document}

\title{Gaussian diagrammatics from Circular Ensembles of random matrices}
\author{Marcel Novaes\\Instituto de F\'isica, Universidade Federal de Uberl\^andia \\Uberl\^andia, MG, 38408-100, Brazil}
\maketitle

\begin{abstract}
We uncover a hidden Gaussian ensemble inside each of the three circular ensembles of random matrices,  providing novel diagrammatic rules for the calculation of moments. The matrices involved are generic complex for $\beta=2$, complex symmetric for $\beta=1$ and complex self-dual for $\beta=4$, and at the last step their dimension must be set to $1-2/\beta$. As an application, we compute moments of traces of submatrices. 
\end{abstract}

\section{Introduction}

The Circular Unitary Ensemble ($CUE$) is the unitary group $U$ equipped with the normalized Haar measure. The Circular Orthogonal Ensemble ($COE$) contains matrices of the form $SS^T$, with $S$ in the CUE, while the Circular Symplectic Ensemble ($CSE$) contains matrices of the form $SS^D$, with $S$ in the CUE. Here $S^T$ and $S^D$ are the transpose and the quaternion dual of $S$. As a result, matrices in the COE are symmetric and matrices in the CSE are self-dual \cite{Mehta} .

Physically, circular ensembles are important as models of random propagators in complex quantum systems, with the particular ensembles corresponding to different symmetry classes (presence or absence of time-reversal and spin rotation invariances) \cite{Haake}. 

Mathematically, circular ensembles can be seen as examples of symmetric spaces, related to the unitary group $U(N)$ itself, in the case of $CUE(N)$, and two of its quotients, namely by the orthogonal group, $U(N)/O(N)$, in the case of $COE(N)$, and by the symplectic group, $U(2N)/Sp(N)$, in the case of $CSE(N)$.

Statistically, eigenvalues of matrices from circular ensembles are perhaps the simplest models of correlated random variables, because they have a constant density and their joint probability distribution consists only of a Vandermonde term, $|\Delta(S)|^{\beta}$, where the Dyson index is $\beta=1$, $2$, $4$ for $COE$, $CUE$, $CSE$, respectively \cite{Dyson}.

These properties of Dyson's circular ensembles can be contrasted, for example, with Wigner's Gaussian ensembles \cite{Mehta,Porter}, introduced in order to model quantum Hamiltonians, for which the spectral density is not constant and the joint probability distribution of eigenvalues contains extra terms besides the Vandermonde.

In this work, we are not interested in spectral statistics but in the complementary problem of the joint distribution of matrix elements. Specifically, the problem of computing moments, i.e. the average value of a product of matrix elements. This can be reduced to the calculation of Weingarten functions \cite{collins}, all of which are known for classical compact Lie groups \cite{CS,CM} and associated symmetric spaces \cite{matsucoe,matsusymm}. 

Our contribution is to uncover a hidden Gaussian ensemble inside each of the three circular ensembles. This allows the introduction of Gaussian diagrammatics, i.e. topological expansions in terms of ribbon graphs, for the calculation of moments of $C\beta E(N)$ in the form of a series in inverse powers of the quantity $\frac{\beta}{2}(N-1)+1$, i.e. in inverse powers of $N+1$, $N$, $2N-1$ for $COE(N)$, $CUE(N)$, $CSE(N)$, respectively. The unusual peculiarity is that in these Gaussian ensembles $N$ appears as a parameter, and they do not correspond to matrices of positive integer dimension. Instead, the dimension should be taken as $1-2/\beta$.

Our results provide new expansions for the Weingarten functions of the three circular ensembles and, indirectly, for the classical compact Lie groups. Other versions of these expansions already have been studied \cite{ramanujan,berko,hciz,expansion,b,novak}, always relating the coefficients with the solution of some combinatorial problem such as counting maps or factorizations of permutations. A direct proof of the equivalence between the present results and previous ones is an open problem.

This paper is organized as follows. In Section 2 we present a short review of the combinatorics of Gaussian matrix models, emphasizing the case involving complex symmetric matrices. In Section 3 we uncover the Gaussian model inside circular ensembles and show thy can be used to obtain moments. In Section 4 we compare our diagrammatics with another diagrammatics for circular ensembles, developed in the context of quantum chaos. In Section 5 we compute, as an application of our approach, statistics of traces of submatrices of dimension $M$ within $COE(N)$. We conclude in Section 6.

\section{Review of Gaussian matrix combinatorics}

Let $0_d$ and $1_d$ denote the $d$-dimensional null matrix and identity matrix, respectively. Let $A^T$ and $\overline{A}$ be the transpose and the complex conjugate of $A$, and $A^\dagger =\overline{A^T}$. With $J=\begin{pmatrix}0_d&1_d\\-1_d&0_d\end{pmatrix}$, define the dual $A^D=JA^TJ^T$.

Let $Z$ denote complex matrices of dimension $d$, with no symmetry when $\beta=2$, symmetric ($Z^T=Z$) when $\beta=1$, and self-dual ($Z^D=Z$) when $\beta=4$. Let 
\be P_\beta(Z)=\frac{1}{\mathcal{G}(d,\beta)}e^{-\Omega_\beta\tr(ZZ^\dagger)}\ee
be the probability distribution of $Z$, with 
\be \mathcal{G}(d,\beta)=\int dZ e^{-\Omega_\beta\tr(ZZ^\dagger)}.\ee

According to Wick's theorem, the integral of a product of $2n$ matrix elements, $n$ from $Z$ and $n$ from $\overline{Z}$, will be given as a sum over all possible pairings between $Z$ and $\overline{Z}$ elements. For $\beta=2$, this means
\be  \int dZ P_2(Z)\prod_{k=1}^n z_{i_{k}j_{k}}\overline{z}_{a_{k}b_{k}}= \sum_{\sigma\in S_n}\prod_{k=1}^n \int dZ P_2(Z) z_{i_{k}j_{k}}\overline{z}_{a_{\pi(k)}b_{\pi(k)}},\ee where $S_n$ is the permutation group. For $\beta=1$ matrices are symmetric, so the pairing is allowed to reverse indices. This leads to  
\be  \int dZ P_1(Z)\prod_{k=1}^{n} z_{i_{2k-1}i_{2k}}\overline{z}_{j_{2k-1}j_{2k}}= \sum_{\sigma\in H_n}\prod_{k=1}^n \int dZ P_1(Z) z_{i_{2k-1}i_{2k}}\overline{z}_{j_{\sigma(2k-1)}j_{\sigma(2k)}},\ee where $H_n$ is the hyperoctahedral group, the wreath product $S_2\wr S_n$. The situation for $\beta=4$ is more convoluted, but it has long been known \cite{osp,cvi,mulase} that models with $\beta=4$ are dual to models with $\beta=1$. Therefore, in what follows we avoid $\beta=4$ for simplicity of exposition. For $\beta=1, 2$ the matrix elements on the diagonal or above it are independent, and the basic covariances are given by 
\be \int dZ P_2(Z) z_{ij}\bar{z}_{km}=\frac{1}{\Omega_2}\delta_{ik}\delta_{jm},\ee and
\be \int dZ P_1(Z) z_{ij}\bar{z}_{km}=\frac{1}{2\Omega_1}(\delta_{ik}\delta_{jm}+\delta_{im}\delta_{jk}).\ee

Wick's rule leads to an elegant diagrammatical formulation of integrals. Matrix elements coming from traces like ${\rm Tr}(ZZ^\dagger)^{k}$ are arranged around vertices, and calculation of basic covariances are represented by edges. The sum over pairings becomes a sum over diagrams or maps. Because every edge must involve one $z$ and one $\overline{z}$, we say they are directed, and by convention they go ``from'' $z$ ``to'' $\overline{z}$. When $Z$ is symmetric, edges may be twisted, resulting in a map that cannot be embedded in an orientable space. 

This kind of combinatorics has been extensively investigated \cite{matrix1,matrix2,matrix3,matrix4,matrix5}. Complex hermitian matrices (Gaussian Unitary Ensemble) give rise to orientable maps, but with undirected edges; generic complex matrices (Complex Ginibre Ensemble) involve orientable and directed maps, which moreover are face-bicolored. For real matrices the maps are neither directed nor orientable, while face-bicoloring holds if the matrices are generic (Real Ginibre Ensemble) and does not hold if the matrices are symmetric (Gaussian Orthogonal Ensemble). 

The model we presently consider for $\beta=1$, with complex symmetric matrices, does not seem to have attracted any attention in this context. It involves directed, non-orientable maps without face-bicoloring.

Let 
\be Z_{\vec{i}\vec{j}}=\prod_{k=1}^{2n} z_{i_kj_k}, \quad Z_{\vec{i}}=\prod_{k=1}^n z_{i_{2k-1}i_{2k}}.\ee The integrals we are interested in are 
\be\label{I2} I_2(\vec{i},\vec{j},\vec{a},\vec{b},n)=\int dZ P_2(Z)e^{-\Omega_2\sum_{q=2}^\infty \frac{1}{q}\tr(ZZ^\dagger)^q}Z_{\vec{i}\vec{j}}\overline{Z}_{\vec{a}\vec{b}},\ee
and
\be\label{I1} I_1(\vec{i},\vec{j},n)=\int dZ P_1(Z)e^{-\Omega_1\sum_{q=2}^\infty \frac{1}{q}\tr(ZZ^\dagger)^q}Z_{\vec{i}}\overline{Z}_{\vec{j}}.\ee
Diagrammatics arises when the exponential is expanded in a Taylor series, 
\be e^{-\Omega_\beta\sum_{q=2}^\infty \frac{1}{q}\tr(ZZ^\dagger)^q}=\sum_{\lambda}\frac{(-\Omega_\beta)^{\ell(\lambda)}}{z_\lambda}\prod_{i=1}^{\ell(\lambda)}{\rm Tr}(ZZ^\dagger)^{\lambda_i},\ee
where the sum is over integer partitions $\lambda=(\lambda_1,\lambda_2,\ldots)$ with no part equal to $1$. The quantity $z_\lambda$ equals $\prod_i \lambda_i \widehat{\lambda_i}!$, where $\widehat{\lambda_i}$ is the multiplicity of $i$ in $\lambda$.

The quantity 
\be p_\lambda(ZZ^\dagger)=\prod_{i=1}^{\ell(\lambda)}{\rm Tr}(ZZ^\dagger)^{\lambda_i}\ee is a power sum symmetric polynomial in the eigenvalues of $ZZ^\dagger$. With this notation, we have 
\be I_\beta=\sum_{\lambda}\frac{(-\Omega_\beta)^{\ell(\lambda)}}{z_\lambda}J_{\beta,\lambda},\ee
with 
\be\label{J2} J_{2,\lambda}(\vec{i},\vec{j},\vec{a},\vec{b},n)=\int dZ P_2(Z)p_\lambda(ZZ^\dagger)Z_{\vec{i}\vec{j}}\overline{Z}_{\vec{a}\vec{b}},\ee
and
\be\label{J1} J_{1,\lambda}(\vec{i},\vec{j},n)=\int dZ P_1(Z)p_\lambda(ZZ^\dagger)Z_{\vec{i}}\overline{Z}_{\vec{j}}.\ee

Using Wick's rule, this leads to a diagrammatic formulation for the integrals, with diagrammatic rules that are a multiplicative combination of weights as follows: The matrix elements in the integrand are vertices of valence 1; A trace $\tr(ZZ^\dagger)^q$ is a vertex of valence $2q$; To every vertex of valence $2q$ we associate a factor $(-\Omega_\beta)$; Every Wick contraction of a $z$ with a $\overline{z}$ is a directed edge; To every edge we associate a factor $(2\Omega_\beta/\beta)^{-1}$; Edges may be twisted if $\beta=1$; To every closed cycle we associate a factor $d$. Diagrams need not be connected. 

Notice that, due to the last rule, the contribution of a diagram is proportional to $d^c$, with $c$ the number of closed cycles it contains. A diagram coming from a certain $\lambda$ gives a contribution proportional to $\Omega_\beta^{-(n+|\lambda|-\ell(\lambda))}$. The quantity $|\lambda|-\ell(\lambda)=r(\lambda)$ is called the rank of the partition $\lambda$.

\subsection{Examples}
Let us consider as an example $n=1$ and $I_1([i_1,i_2],[j_1,j_2],1)$. The simplest diagrams have no vertices of even valence ($\lambda$ being the empty partition), and consist of a single edge connecting the two vertices of valence one, as in Figure 1. There are two of these, one taking $i_1$ to $j_1$ and the other being twisted in order to take $i_1$ to $j_2$. Together they contribute
\be J_{1,\emptyset}([i_1,i_2],[j_1,j_2],1)=\frac{1}{2\Omega_1}(\delta_{i_1j_1}\delta_{i_2j_2}+\delta_{i_1j_2}\delta_{i_2j_1}).\ee

\begin{figure*}[t]
\includegraphics[scale = 0.55 ]{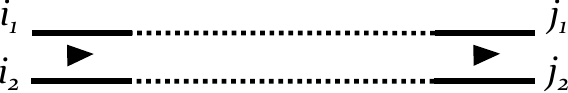}\hspace{0.1cm}
\caption{Leading order diagram for $I_1([i_1,i_2],[j_1,j_2])$, corresponding to the empty partition. This diagram is proportional to $\delta_{i_1j_1}\delta_{i_2j_2}$.}
\end{figure*}

Since $\lambda$ cannot have parts equal to $1$, the next simple case is $\lambda=(2)$. Using a computer algebra system we find that
\be  J_{1,(2)}([i_1,i_2],[j_1,j_2],1)=\frac{(-\Omega_1)}{(2\Omega_1)^3}(2d^3+4d^2+10d+8)(\delta_{i_1j_1}\delta_{i_2j_2}+\delta_{i_1j_2}\delta_{i_2j_1}),\ee 
meaning that, containing one vertex of valence $4$, there are: $2$ diagrams with three closed cyles, $4$ diagrams with two closed cycles, 10 diagrams with one closed cycle and 8 diagrams with no closed cycles. We show four of these diagrams in Figure 2. Dashed lines represent Wick connections. Diagram a) has no closed cycles, diagram b) has one closed cycle, diagram c) has two closed cycles, diagram d) has three closed cycles. Diagrams c) and d) are not connected.

\begin{figure*}[t]
\includegraphics[scale = 0.55]{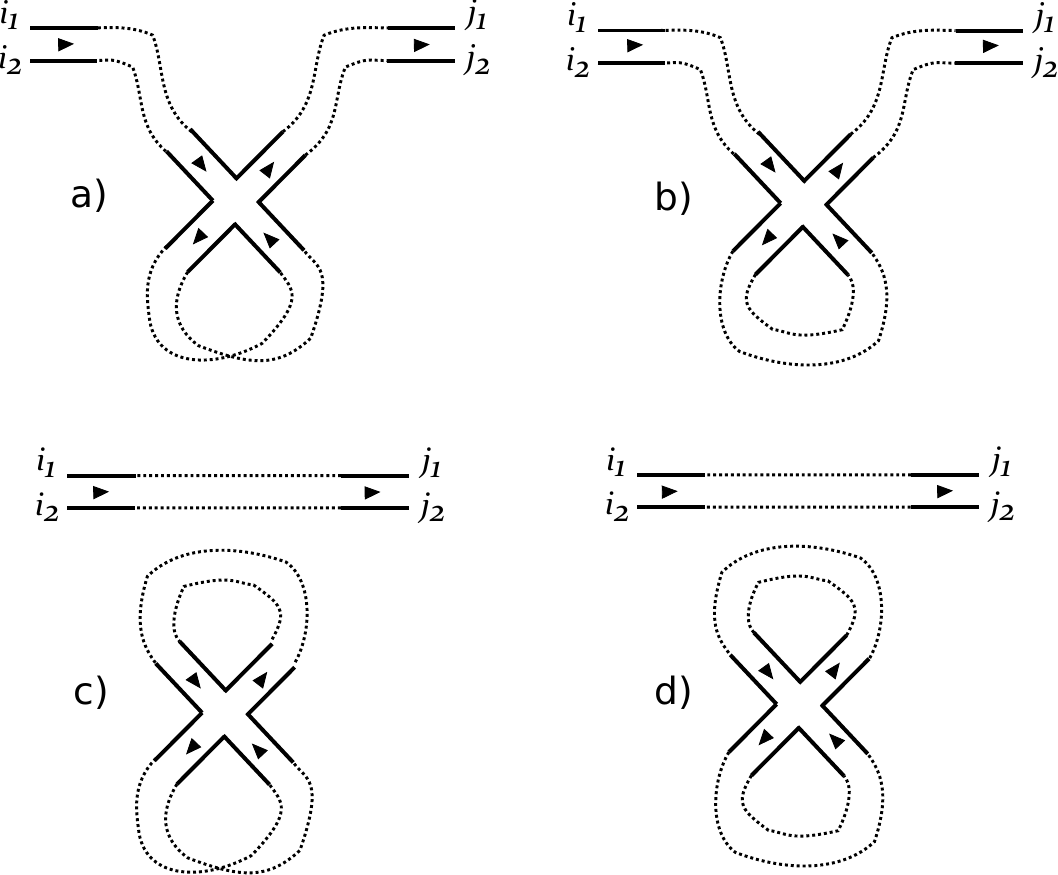}\hspace{0.1cm}
\caption{Four diagrams for $J_{1,(2)}([i_1,i_2],[j_1,j_2])$, with different numbers of closed cycles. Two of them are orientable, two are non-orientable, and all proportional to $\delta_{i_1j_1}\delta_{i_2j_2}$. Small black arrows denote the directionality of the edges.}
\end{figure*}

A little reflection shows that, for a general partition, 
\be J_{1,\lambda}([i_1,i_2],[j_1,j_2],1)=\frac{(-\Omega_1)^{\ell(\lambda)}}{(2\Omega_1)^{|\lambda|+1}}j_{1,\lambda}(d)(\delta_{i_1j_1}\delta_{i_2j_2}+\delta_{i_1j_2}\delta_{i_2j_1}),\ee
where $j_{1,\lambda}(d)$ is a polynomial in the dimension $d$. The first such polynomials are, for partitions of rank 2:
\be j_{1,(3)}=5d^4+ 16d^3+49d^2 +74d +48,\ee
\be j_{1,(2,2)}=4d^6+ 16d^5+92d^4 +224d^3 +412d^2+ 688d+ 384;\ee
and, for partitions of rank 3:
\be j_{1,(4)}=14d^5+ 64d^4+242d^3 +528d^2+ 688d+ 384,\ee
\be j_{1,(3,2)}=10d^7+ 52d^6+356d^5 +1180d^4 +3410d^3+ 6568d^2+ 7624d+3840.\ee
\be j_{1,(2,2,2)}=8d^9+48d^8+432d^7+1744d^6+7704d^5+21568d^4+52912d^3+92992d^2+99072d+46080.\ee

We notice that $j_{1,\lambda}(d)$ is of degree $d^{|\lambda|+\ell(\lambda)}$, and the coefficient of the largest power is a product of Catalan numbers,
\be  [d^{|\lambda|+\ell(\lambda)}]j_{1,\lambda}=\prod_{i=1}^{\ell(\lambda)}\frac{1}{\lambda_i+1}{2\lambda_i \choose \lambda_i}.\ee The proof of this fact is analogous to the complex hermitian case.

For a more generic example, in Figure 3 we show a diagram that contributes to $J_{1,(4,3)}(\vec{i},\vec{j},3)$. It has one vertex of valence 6 and one vertex of valence 8. It has two closed cycles. Two of its edges are twisted. Its contribution is $\frac{\Omega_1^2d^2}{(2\Omega_1)^{10}}\delta_{i_1j_1}\delta_{i_2j_4}\delta_{i_3j_2}\delta_{i_4j_6}\delta_{i_5j_5}\delta_{i_6j_3}$.

\begin{figure*}[t]
\includegraphics[scale = 0.7]{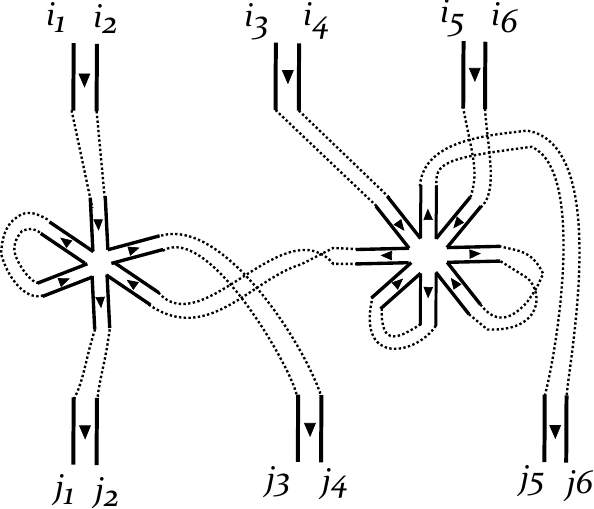}\hspace{0.1cm}
\caption{A diagram for $J_{1,(4,3)}$ at $n=3$, which has two closed cycles, two twisted edges and evaluates to $\frac{\Omega_1^2d^2}{(2\Omega_1)^{10}}\delta_{i_1j_1}\delta_{i_2j_4}\delta_{i_3j_2}\delta_{i_4j_6}\delta_{i_5j_5}\delta_{i_6j_3}$. Small black arrows denote the directionality of the edges.}
\end{figure*}

\section{Moments of Circular Ensembles}

We now show how the Gaussian matrix models of the previous Section are related to the Circular Ensembles. 

Let $S$ be in $C\beta E(N)$ and let $Z$ be its $d\times d$ top left block. The distribution of $Z$ is given by \cite{trunc1,trunc2,trunc3,trunc4}
\be \frac{1}{\mathcal{V}(N,d,\beta)}\det(1-ZZ^\dagger)^{\frac{\beta}{2}(N-2d+1)-1},\ee
with $\mathcal{V}(N,d,\beta)=\int dZ \det(1-ZZ^\dagger)^{\frac{\beta}{2}(N-2d+1)-1}$. This normalization can be computed by changing variables to the real positive eigenvalues of $X=ZZ^\dagger$. The jacobian of this transformation is $|\Delta(X)|^\beta$, which gives
\be \mathcal{V}(N,d,\beta)=A_d\int dX \det(1-X)^{\frac{\beta}{2}(N-2d+1)-1}|\Delta(X)|^\beta,\ee where $A_d$ comes from integration over the eigenvectors. This is an integral of Selberg type \cite{selberg}, whose solution is
\be \mathcal{V}(N,d,\beta)=\prod_{j=1}^d\frac{\Gamma(1+\beta(j-1)/2)\Gamma(\beta(N-d-j+1)/2)\Gamma(1+\beta j/2)}{\Gamma(1+\beta(N-j)/2)\Gamma(1+\beta/2)}.\ee

By definition, as long as all the indices in the matrix elements are smaller than $d$, we can consider them either as elements of $Z$ or of $S$. This means that
\be\label{theu} \frac{1}{\mathcal{V}(N,d,2)}\int dZ \det(1-ZZ^\dagger)^{(N-2d)}Z_{\vec{i}\vec{j}}\overline{Z}_{\vec{a}\vec{b}}=\langle S_{\vec{i}\vec{j}}\overline{S}_{\vec{a}\vec{b}}\rangle_{CUE(N)},\ee
and
\be\label{theo} \frac{1}{\mathcal{V}(N,d,1)}\int dZ \det(1-ZZ^\dagger)^{\frac{1}{2}(N-2d+1)-1}Z_{\vec{i}}\overline{Z}_{\vec{j}}=\langle S_{\vec{i}}\overline{S}_{\vec{j}}\rangle_{COE(N)},\ee and these quantities, after the integrals have been calculated, are actually independent of $d$. 

Using $\det=e^{{\rm Tr}\log}$ we can write, as long as $f$ only involves elements of $S$ inside $Z$,
\be \langle f(S,\overline{S})\rangle_{C\beta E(N)}=\frac{1}{\mathcal{V}(N,d,\beta)}\int dZ e^{-\omega_\beta(d)\tr(ZZ^\dagger)}e^{-\omega_\beta(d)\sum_{q=2}^\infty \frac{1}{q}\tr(ZZ^\dagger)^q}f(Z,\overline{Z}),\ee 
with $\omega_\beta(d)=\frac{\beta}{2}(N-2d+1)-1$.

This looks very similar to the Gaussian models, Eqs. (\ref{I2}) and (\ref{I1}), but not quite identical, in particular because the normalizations are different, $\mathcal{V}(N,d,\beta)\neq\mathcal{G}(d,\beta)$. The Gaussian normalization is 
\be \mathcal{G}(d,\beta)=A_d\int dX e^{-\Omega_\beta{\rm Tr}X}|\Delta(X)|^\beta,\ee
which equals
\be \mathcal{G}(d,\beta)=A_d\Omega_\beta^{-\beta d(d-1)/2-d}\prod_{j=1}^d\frac{\Gamma(1+\beta(j-1)/2)\Gamma(1+\beta j/2)}{\Gamma(1+\beta/2)}.\ee So
\be \frac{\mathcal{V}(N,d,\beta)}{ \mathcal{G}(d,\beta)}=\Omega_\beta^{\beta d(d-1)/2+d}\prod_{j=1}^d\frac{\Gamma(\beta(N-d-j+1)/2)}{\Gamma(1+\beta(N-j)/2)}.\ee

Now, the crucial observation is this: if we let
\be\label{d} d\to 1-2/\beta,\ee
we get 
\be \frac{\mathcal{V}(N,d,\beta)}{ \mathcal{G}(d,\beta)}\to \Omega_\beta^0\prod_{j=1}^d1=1,\ee
and then we do arrive at the Gaussian model, with 
\be \Omega_\beta=\omega_\beta(1-2/\beta)=\frac{\beta}{2}(N-1)+1.\ee

Therefore, we conclude that moments of circular ensembles can be computed using the diagrammatic rules associated with Gaussian models, given in Eqs. (\ref{J2}) and (\ref{J1}), provided we use $\Omega_\beta$ as above. 

The map (\ref{d}) gives $d=-1$, $0$, $1/2$ for $\beta=1$, $2$, $4$, respectively. These are not positive integers as one would expect of a dimension. But these values must be used only after the result has been computed for formal $d$ and found as a polynomial in $d$.

That this approach works for the $CUE$ has already been discussed in the physics context \cite{qc1,qc2,qc3,qc4}. In that case, taking $d\to 0$ rules out the presence of closed cycles in the diagrams. That does not happen for the $COE$, in which case the theory prescribes that the contribution of a diagram with vertex structure $\lambda$ and $c$ closed cycles is proportional to
\be \frac{1}{z_\lambda}\left(\frac{-1}{2}\right)^{\ell(\lambda)}\frac{(-1)^c}{(N+1)^{|\lambda|-\ell(\lambda)+n}}.\ee

\subsection{Example}

Let us take again the simplest example, with $n=1$,
\be \langle S_{i_1i_2}\overline{S}_{j_1j_2}\rangle_{COE(N)}=\frac{\delta_{i_1j_1}\delta_{i_2j_2}+\delta_{i_1j_2}\delta_{i_2j_1}}{N+1}.\ee

The trivial diagram with empty $\lambda$ in the Gaussian model, Figure 1, already agrees with the exact result. Therefore, all other diagrams must cancel out and give a vanishing overall contribution, something which is not at all trivial. 

But indeed, we do have $ j_{1,(2)}(d=-1)=0,$ so the first correction vanishes. 

We have $j_{1,(3)}(-1)=12$, and $j_{1,(2,2)}(-1)=64$. Taking into account the factor $\frac{1}{z_\lambda}\left(\frac{-1}{2}\right)^{\ell(\lambda)}$ we get
\be -\frac{12}{6}+\frac{64}{12}=0,\ee so the second correction vanishes. And we have $j_{1,(4)}(-1)=32,$ $j_{1,(3,2)}(-1)=240$ and $j_{1,(2,2,2)}(-1)=2304,$ which leads to a vanishing third correction: 
\be -\frac{32}{8}+\frac{240}{24}-\frac{2304}{384}=0.\ee

\section{The semiclassical diagrammatics}

Curiously, another diagrammatical formulation of moments in $C\beta E(N)$ already exists. It was developed by physicists working with semiclassical path integral approximations in quantum chaos \cite{qc1,qc2}. This was indeed the original motivation for the present work. 

The semiclassical diagrammatics is as follows: any given moment is written as a sum over diagrams, with weights of $-N$ for each vertex of valence larger than one, $1/N$ for each edge, with orientability being required when $\beta=2$. Closed cycles are not allowed. 

\begin{figure*}[t]
\includegraphics[scale = 0.6]{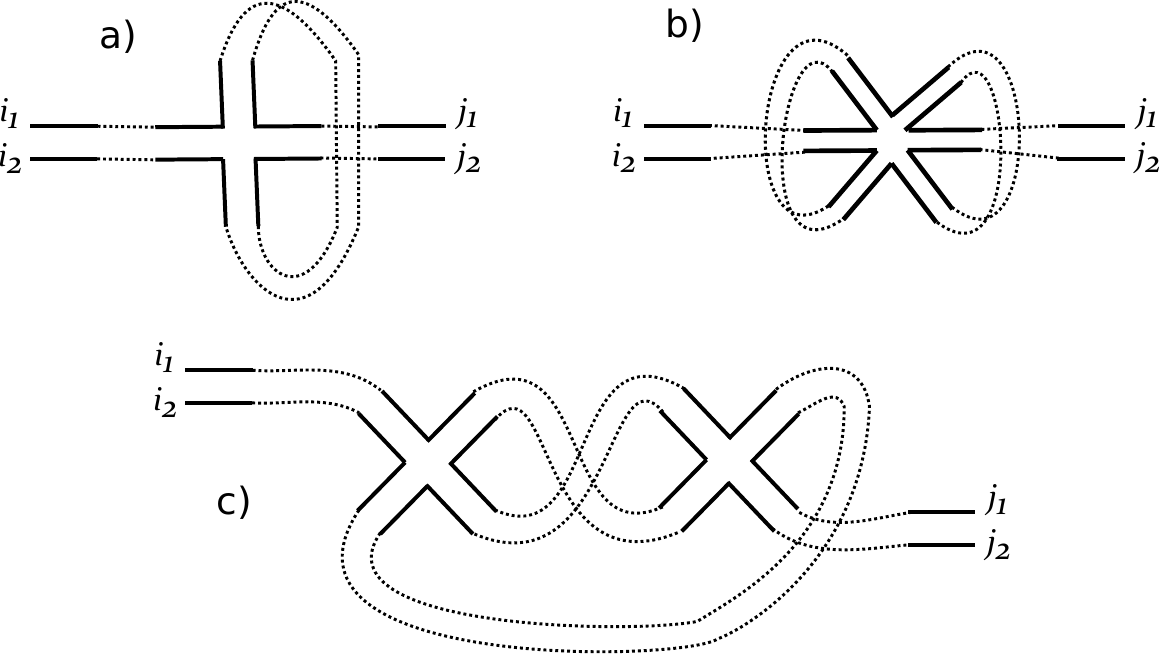}
\caption{Diagrams for the diagrammatics of $\langle S_{i_1i_2}\overline{S}_{j_1j_2}\rangle_{COE(N)}$ according to the semiclassical model. Edges are undirected, and the weights of edges and vertices are different than in Figures 1-3. These diagrams reproduce the first leading orders of an infinite series in $1/N$.}
\end{figure*}

These rules are exactly the ones we obtained, when $\beta=2$. So our Gaussian model coincides with the semiclassical model of quantum chaos. However, this is not so for $\beta=1$. There are two differences in the rules themselves, because in our model the weights are $-(N+1)/2$ for each vertex of valence larger than one and $1/(N+1)$ for each edge. There is also another difference, which is that in our model the edges are directed, while this does not hold in the semiclassical model. 

For comparison, let us sketch the calculation of $\langle S_{i_1i_2}\overline{S}_{j_1j_2}\rangle_{COE(N)}$ according to the semiclassical model. The leading order is given by the diagrams in Figure 1, except now they contribute $1/N$. Then there are diagrams like the ones in Figure 4. Their contributions are: a) $-1/N^2$ (one vertex of valence 4); b) $-1/N^3$ (one vertex of valence 6); c)  $1/N^3$ (two vertices of valence 4). Indeed, we have just arrived at the first three terms in the $1/N$ expansion of $1/(N+1)$. When all possible diagrams are summed, the exact result is recovered in the form of the geometric series.

Notice how in the diagrammatics obtained in the present work, the same calculation requires a single diagram, with the contribution from all others being zero due to nontrivial cancellations, just like happens for $CUE(N)$. In contrast, in the semiclassical diagrammatics it is the opposite: infinitely many diagrams are required. 

Actually, the semiclassical diagrammatics for $\beta=1$ can be implemented using matrix integrals involving non-symmetric real matrices \cite{AP}. This is related to the fact that the Weingarten function of $COE(N)$ is actually equal to the Weingarten function of the real orthogonal group in dimension one higher, $O(N+1)$. This equality was established by Matsumoto, who pointed out that it must be grounded in the fact that $COE(N)\sim U(N)/O(N)$, but still remarked that it was ``quite mysterious''. A Gaussian model for the combinatorics of orthogonal group moments in terms of real matrices was indeed derived in \cite{expansion}.

\section{Application: traces of submatrices}

In \cite{traces}, Jiang and Matsumoto studied the traces of matrices from the $COE(N)$. In particular, they showed that 
\be \lim_{N\to \infty}\langle |p_\lambda(Z)|^2\rangle_{COE(N)}=2^{\ell(\lambda)}z_\lambda.\ee
They actually derive an exact formula for the above quantity, but it is not quite explicit because it depends on Jack characters, the coefficients in the expansion of power sums into Jack polynomials. They also show that $\langle p_\lambda(Z)p_\mu(\overline{Z})\rangle_{COE(N)}$ decays like $1/N$ if $|\mu|=|\lambda|$ but $\mu\neq \lambda$.

As an application of the results we have obtained, we compute averages of traces of a submatrix $Z$ of dimension $M$ inside a $COE(N)$ matrix. Results from \cite{traces} then correspond to the particular case $M=N$. 

Let $B(M,N)$ denote the $M\times M$ upper left block of $COE(N)$ matrices. When we wish to compute something like $\langle p_\lambda(Z)p_\mu(\overline{Z})\rangle_{B(M,N)}$ using our Gaussian model, we write 
\be \langle p_\lambda(Z)p_\mu(\overline{Z})\rangle_{B(M,N)}=\sum_{i_1,...,i_n=1}^M\sum_{j_1,...,j_n=1}^M\left\langle\prod_{k=1}^n  Z_{i_k,i_{\pi(k)}}\overline{Z}_{j_k,j_{\sigma(k)}}\right\rangle_{COE(N)},\ee where $\pi,\sigma$ are any permutations with cycle type $\lambda,\mu$. Then, we use the diagrammatic rules of Section 3 to compute the average, and finally we sum over $\vec{i}$ and $\vec{j}$. In this way, closed cycles associated with $\vec{i},\vec{j}$ indices have a weight of $M$, while closed cycles within the average have weight $(-1)$. This calculation is carried out in a computer algebra system, so only the simplest partitions can be addressed.

Averages of traces of submatrices were studied for the unitary group in \cite{novaku}, and averages of their Schur polynomials appear in \cite{remark}. These approaches rely on the usual characters of the permutation group, which are available in computer algebra systems, so we do not address the $\beta=2$ case.

The simplest average,
\be \langle |p_{(1)}(Z)|^2\rangle_{B(M,N)}=\frac{2M}{N+1},\ee
is actually exact and easy to derive. For $n=2$ we obtain two corrections to the leading order results,  
\be \langle |p_{(2)}(Z)|^2\rangle_{B(M,N)}=\frac{4M(M+1)}{(N+1)^2}-\frac{4M(M+3)}{(N+1)^3}+\frac{2M(M+15)}{(N+1)^4}+\cdots\ee
and
\be \langle |p_{(1,1)}(Z)|^2\rangle_{B(M,N)}=\frac{8M^2}{(N+1)^2}-\frac{16M}{(N+1)^3}-\frac{8M(3M-7)}{(N+1)^4}+\cdots.\ee 

Notice that the large-$N$ asymptotics may be rather rich. $ \langle |p_{(2)}(Z)|^2\rangle$ equals, to leading order, $4M(M+1)/N^2+O(N^{-3})$ if $M$ is held fixed, $4-10/N^2+O(N^{-3})$ if $M=N$ and $4\xi^2+4\xi(1-\xi)/N+O(N^{-2})$ if $M=\xi N$ with $0<\xi<1$.

In constrast with the $CUE$, quantities like $\langle p_\lambda(Z)p_\mu(\overline{Z})\rangle_{COE(N)}$ do not necessarily vanish when $\mu\neq \lambda$. For example,
\be \langle p_{(2)}(Z)p_{(1,1)}(\overline{Z})\rangle_{B(M,N)}=\frac{8M}{(N+1)^2}-\frac{8M(M+1)}{(N+1)^3}+\frac{4M(7M+1)}{(N+1)^4}+\cdots.\ee Again, the asymptotical behavior depends on how $M$ related to $N$. The above quantity becomes $8M/N^2+O(N^{-3})$ if $M$ is held fixed, $20/N^2+O(N^{-3})$ if $M=N$ and $8\xi(1-\xi)/N+O(N^{-2})$ if $M=\xi N$ with $0<\xi<1$.

When $n=3$ we obtain one correction to the leading order results:
\be \langle |p_{(3)}(Z)|^2\rangle_{B(M,N)}=\frac{6M(M^2+3M+4)}{(N+1)^3}-\frac{18M(M^2+7M+8)}{(N+1)^4}+\cdots,\ee
\be \langle |p_{(2,1)}(Z)|^2\rangle_{B(M,N)}=\frac{8M(M^2+M+4)}{(N+1)^3}-\frac{8M(M^2+19M+16)}{(N+1)^4}+\cdots,\ee
and
\be \langle |p_{(1,1,1)}(Z)|^2\rangle_{B(M,N)}=\frac{48M^3}{(N+1)^3}-\frac{288M^2}{(N+1)^4}+\cdots.\ee
The off-diagonal covariances in this case are
\be \langle p_{(3)}(Z)p_{(2,1)}(\overline{Z})\rangle_{B(M,N)}=\frac{24M(M+1)}{(N+1)^3}-\frac{24M(M^2+4M+7)}{(N+1)^4}+\cdots,\ee
\be \langle p_{(3)}(Z)p_{(1,1,1)}(\overline{Z})\rangle_{B(M,N)}=\frac{48M}{(N+1)^3}-\frac{144M(M+1)}{(N+1)^4}+\cdots,\ee
and
\be \langle p_{(2,1)}(Z)p_{(1,1,1)}(\overline{Z})\rangle_{B(M,N)}=\frac{48M^2}{(N+1)^3}-\frac{48M(M^2+M+4)}{(N+1)^4}+\cdots.\ee
When $M=N$, the leading order of these results coincide with those from \cite{traces}.

\section{Conclusion}

We have found Gaussian matrix models that provide novel diagrammatic rules for the calculation of moments in circular ensembles, $C\beta E(N)$. The matrices involved are generic complex for $\beta=2$, complex symmetric for $\beta=1$ and complex self-dual for $\beta=4$, and their dimension must be set to $1-2/\beta$. The result is an expansion in inverse powers of $N$ for $\beta=2$, of $N+1$ for $\beta=1$ and of $2N-1$ for $\beta=4$.

The expansion parameters $N+1$ and $2N-1$ for $\beta=1,4$ must somehow be related to the Weingarten function identities 
\be\label{iden} {\rm Wg}^{COE(N)}={\rm Wg}^{O(N+1)},\quad {\rm Wg}^{CSE(N)}={\rm Wg}^{Sp(2N-1)},\ee and in fact our results lead to new diagrammatic expansions in inverse powers of $N$ for the moments of  $O(N)$ and $Sp(N)$. This might shed some light on the above identities, which at the moment are still little more than coincidences.

The complex symmetric matrix model we developed here for $COE(N)$ must be equivalent to the generic real matrix model developed in \cite{expansion} for $O(N+1)$ and the corresponding semiclassical model \cite{AP}. However, this equality between very different matrix models is not obvious a priori and a direct proof would be quite interesting.


\begin{thebibliography}{99}

\bibitem{Mehta} M. L. Mehta, Random matrices (Elsevier, 2004).

\bibitem{Haake} F. Haake, S. Gnutzmann, M. Ku\'s, Quantum Signatures of Chaos (Springer, 2019).

\bibitem{Dyson} F. J. Dyson, Statistical Theory of the Energy Levels of Complex Systems. I. J. Math. Phys. 3, 140 (1962); II. J. Math. Phys. 3, 157 (1962); III. J. Math. Phys. 3, 166 (1962).

\bibitem{Porter} C. E. Porter (Ed.), Statistical Theories of Spectra: Fluctuations (Academic Press, 1965).

\bibitem{collins} B. Collins, Moments and cumulants of polynomial random variables on unitary groups, the Itzykson-Zuber integral, and free probability. Int. Math. Res. Notices 2003, 953 (2003).

\bibitem{CS} B. Collins and P. \'Sniady, Integration with respect to the Haar measure on unitary, orthogonal and symplectic group. Commun. Math. Phys. 264, 773 (2006).

\bibitem{CM} B. Collins and S. Matsumoto, On some properties of orthogonal Weingarten functions
J. Math. Phys. 50, 113516 (2009).

\bibitem{matsucoe} S. Matsumoto, General moments of matrix elements
from circular orthogonal ensembles. Random Matrices: Theory and Applications 1, 1250005 (2012).

\bibitem{matsusymm} S. Matsumoto, Weingarten calculus for matrix ensembles associated with compact symmetric spaces. Random Matrices: Theory and Applications 2, 1350001 (2013).

\bibitem{ramanujan} S. Matsumoto, Jucys–Murphy elements, orthogonal matrix integrals,
and Jack measures. Ramanujan J 26, 69 (2011).

\bibitem{berko} G. Berkolaiko and J. Kuipers, Combinatorial theory of the semiclassical evaluation of transport moments II: Algorithmic approach for moment generating functions. J. Math. Phys. 54, 123505 (2013);

\bibitem{hciz}  I. Goulden, M. Guay-Paquet, J. Novak, Monotone Hurwitz Numbers and the HCIZ Integral. Annales mathématiques Blaise Pascal 21, 71 (2014).

\bibitem{expansion} M. Novaes, Expansion of polynomial Lie group integrals in terms of certain maps on surfaces, and factorizations of permutations. J. Phys. A: Math. Theor. 50, 075201 (2017).

\bibitem{b} V. Bonzom, G. Chapuy, M. Do{\l}ega, $b$-monotone Hurwitz numbers: Virasoro constraints, BKP hierarchy, and $O(N)$-BGW integral. arXiv preprint arXiv:2109.01499, 2021.

\bibitem{novak} S. Matsumoto and J. Novak, Jucys–Murphy elements and unitary matrix integrals. Int. Math. Res. Not. 2 362 (2013).

\bibitem{osp} R.L. Mkrtchian, The Equivalence of $Sp(2N)$ and $SO(-2N)$ gauge theories. Phys. Lett. B 105, 174 (1981).

\bibitem{cvi} P. Cvitanovic abd A.D. Kennedy, Spinors in negative dimensions. Phys. Scripta 26, 5 (1982).

\bibitem{mulase} M. Mulase and A. Waldron, Duality of Orthogonal and Symplectic Matrix Integrals and Quaternionic Feynman Graphs. Commun. Math. Phys. 240, 553 (2003).

\bibitem{matrix1} G. ’t Hooft, A planar diagram theory for strong interactions. Nucl. Phys. B 72, 461 (1974).

\bibitem{matrix2} D. Bessis, C. Itzykson and J.B. Zuber, Quantum field theory techniques in graphical enumeration. Adv. Appl. Math. 1, 109 (1980).

\bibitem{matrix3} P. Di Francesco, P. Ginsparg and J. Zinn-Justin, 2D gravity and random matrices, Phys. Rep. 254, 1 (1995).

\bibitem{matrix4} T.R. Morris, Chequered surfaces and complex matrices. Nucl. Phys. B 356, 703 (1991).

\bibitem{matrix5} I. Goulden and D. Jackson, Maps in locally orientable surfaces and integrals over real symmetric surfaces. Canadian J. Math., 49, 865 (1997).

\bibitem{trunc1} H.U. Baranger and P.A. Mello, Mesoscopic transport through chaotic cavities: a random $S$-matrix theory approach. Phys. Rev. Lett. 73, 142 (1994).

\bibitem{trunc2} J.-L. Pichard, R.A. Jalabert and C.W.J. Beenakker, Universal signatures of chaos in ballistic transport. Europhys. Lett. 27, 255 (1994).

\bibitem{trunc3} C.W.J. Beenakker, Random-matrix theory of quantum transport Rev. Mod. Phys. 69, 731 (1997).

\bibitem{trunc4} P.J. Forrester, Quantum conductance problems and the Jacobi
ensemble. J. Phys. A: Math. Gen. 39, 6861 (2006).

\bibitem{selberg} P. J. Forrester and O. S. Warnaar, The importance of the Selberg
integral, Bull. Am. Math. Soc. 45, 489 (2008).

\bibitem{qc1} S. Heusler, S. M\"uller, P. Braun and F. Haake, Semiclassical theory of chaotic conductors. Phys. Rev. Lett. 96, 066804 (2006).
 
\bibitem{qc2} S. M\"uller, S. Heusler, P. Braun and F. Haake, Semiclassical approach to chaotic quantum transport. New J. Phys. 9, 12 (2007).

\bibitem{qc3} G. Berkolaiko and J. Kuipers, Universality in chaotic quantum transport: the concordance between random-matrix and semiclassical theories Phys. Rev. E 85, 045201 (2012).

\bibitem{qc4} G. Berkolaiko and J. Kuipers, Combinatorial theory of the semiclassical evaluation of transport moments. I. Equivalence with the random matrix approach J. Math. Phys. 54, 112103 (2013).

\bibitem{AP} M. Novaes, Semiclassical matrix model for quantum chaotic transport with time-reversal symmetry. Annals of Physics 361, 51 (2015).


\bibitem{traces} T. Jiang and S. Matsumoto, Moments of traces of circular beta-ensembles. The Annals of Probability 43, 3279 (2015).

\bibitem{novaku} J. Novak, Truncations of random unitary matrices
and Young tableaux. Electronic Journal of Combinatorics 14, R21 (2007).


\bibitem{remark} Y.V. Fyodorov and B.A. Khoruzhenko, A few remarks on colour–flavour transformations, truncations of random unitary matrices, Berezin reproducing kernels and Selberg-type integrals. J. Phys. A: Math. Theor. 40, 669 (2007).




\end{thebibliography}
\end{document}